\documentclass{vgtc}                          

\graphicspath{{figures/}{pictures/}{images/}{./}}

\usepackage{times}

\usepackage{tabu}
\usepackage{booktabs}
\usepackage{lipsum}
\usepackage{mwe}
\usepackage{mathptmx}

\onlineid{0}
\vgtccategory{Research}
\vgtcinsertpkg

\title{The Living Library of Trees:\\Mapping Knowledge Ecology in the Arnold Arboretum}

\author{
Johan Malmstedt\\ \scriptsize Gothenburg University %
\thanks{Email: johan.malmstedt@gu.se} %
\and Giacomo Nanni\\ \scriptsize Free University of Berlin %
\thanks{Email: giacomo.nanni@fu-berlin.de} %
\and Dario Rodighiero\\ \scriptsize University of Groningen %
\thanks{Email: d.rodighiero@rug.nl}
}

\teaser{
  \centering
  \includegraphics[width=\linewidth]{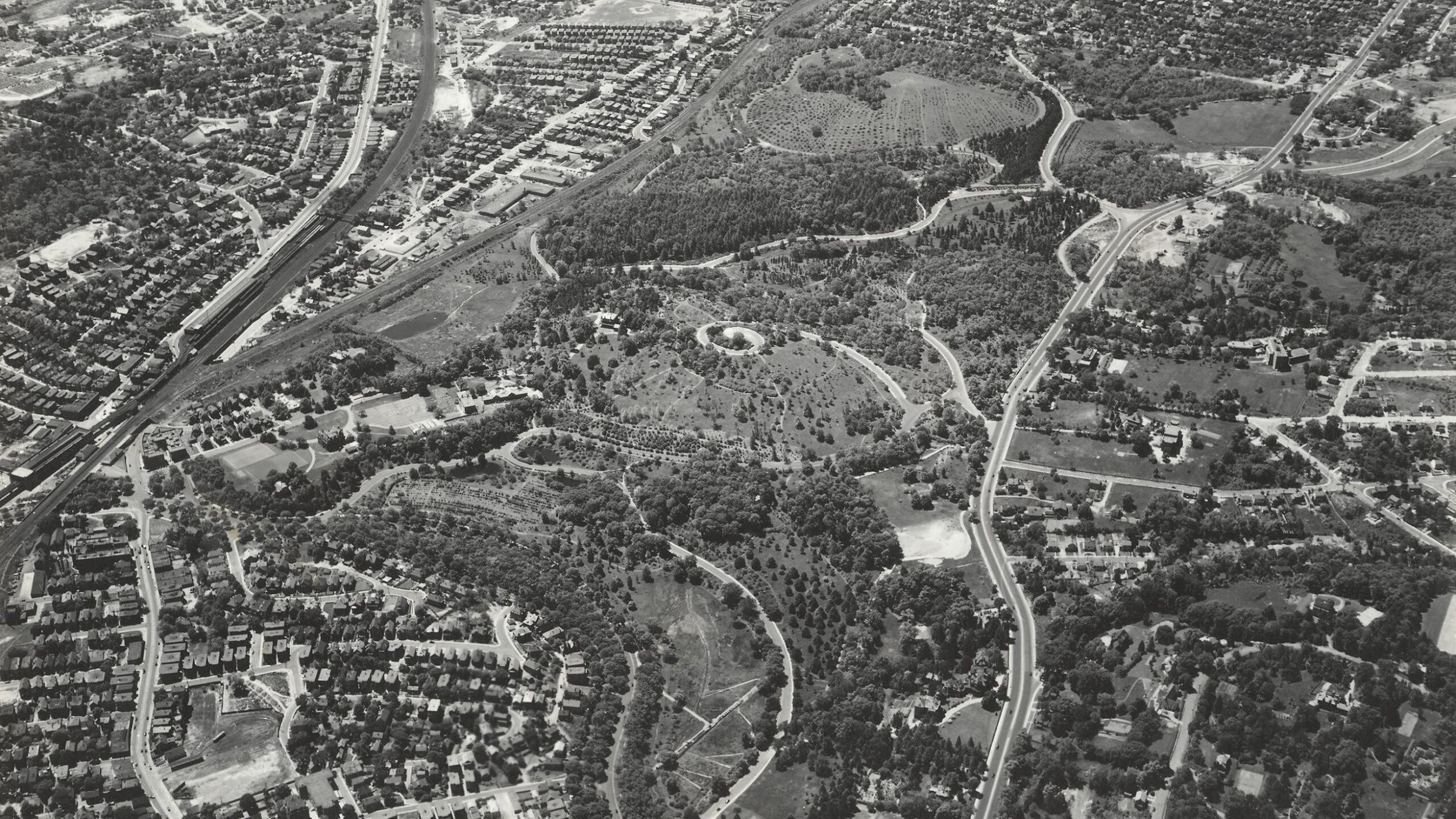}
\caption{An early twentieth-century aerial photograph of the Arnold Arboretum shows a carefully planned mosaic of paths, groves, and open lawns. More than picturesque scenery, it is a working landscape for science and public learning [Courtesy of the Arnold Arboretum of Harvard University].} \label{fig:teaser}
}

\abstract{
As biodiversity loss and climate change accelerate, botanical gardens serve as vital infrastructures for research, education, and conservation. This project focuses on the Arnold Arboretum of Harvard University, a 281-acre living museum founded in 1872 in Boston. Drawing on more than a century of curatorial data, the research combines historical analysis with computational methods to visualize the biographies of plants and people. The resulting platform reveals patterns of care and scientific observations, along with the collective dimensions embedded in botanical data. Using techniques from artificial intelligence, geospatial mapping, and information design, the project frames the arboretum as a system of shared agency—an active archive of more-than-human affinities that records the layered memory of curatorial labor, the situated nature of knowledge production, and the potential of design to bridge archival record and future care.}


\begin{document}


\firstsection{Introduction}

\maketitle

Beyond their role as public spaces, botanical gardens are key sites in responding to biodiversity loss and climate change. They act as biocultural infrastructures, linking public participation with scientific research and bridging policy, education, and aesthetics. The Arnold Arboretum of Harvard University, founded in 1872 and among the oldest public \textit{arboreta} in North America, illustrates this role through curatorial practices that reveal how people and plants adapt to and learn from each other. Yet, despite their importance as information centers, botanical gardens remain understudied as knowledge infrastructures.

This project approaches the arboretum as both a field and a medium—between locality and the herbaceous environment, between data and care~\cite{ref2}, between \textit{science-in-the-making} and living taxonomy~\cite{ref3}. We trace the flows of information sustaining this ``planted catalog''~\cite{ref1} across formats, from nineteenth-century ledgers to contemporary datasets. In this light, the arboretum emerges as a living knowledge system where care, classification, and ecological presence are collectively shaped, and where the archive becomes a lens for reimagining relations between humans and the more-than-human world. Its dual character—as both living collection and historical record—offers a framework for examining how scientific authority is built, how environmental change becomes inscribed, and how institutional memory is mediated by human judgment and nonhuman growth alike.


\section{Research Framework}

This project is situated at the intersection of ethnobotany~\cite{ref4}, digital humanities~\cite{ref5}, and science and technology studies (STS)~\cite{ref6}. Each of these fields offers critical tools for examining how inscriptions---whether botanical, textual, or digital---are produced, circulated, and sustained, particularly in contexts involving both human and nonhuman actors~\cite{ref7}. By approaching the arboretum as a site of knowledge infrastructure, this study proposes a visual framework for exploring not only what knowledge is preserved within a living archive, but also how that knowledge has been, and continues to be, created, maintained, and interpreted over time.

\subsection{Botanical Gardens as Knowledge Infrastructures}

Botanical gardens have long served as sites of classification, collection, and colonial encounter. Scholars such as Tsing~\cite{ref8}, Haraway~\cite{ref9}, and Myers~\cite{ref10} have examined their role in shaping understandings of biodiversity, emphasizing the relational character of ecological knowledge over claims of objectivity. From this perspective, the arboretum functions as a knowledge infrastructure---a system through which the natural world is translated into data and rendered legible to both scientific and public audiences.

While traditional studies of botanical gardens have centered on taxonomic history, more recent research has turned toward the less visible labor of curating living collections. Tuers~\cite{ref1} describes these ``border spaces between the locality and the herbarium'' as zones where plants, metadata, and curatorship converge---crucial for understanding how botanical data takes form. In articulating the concept of the ``planted catalog,'' Tuers argues that gardens do not merely reflect nature; they shape how it is preserved, retrieved, and reimagined. With its extensive physical and digital archives, the Arnold Arboretum offers a uniquely rich site for exploring these dynamics across historical and contemporary registers.

\subsection{Interspecies Affinity and the Politics of Care}

This study adopts the concept of interspecies affinity---introduced by Haraway~\cite{ref9} and further developed by scholars such as John Hartigan Jr.~\cite{hartigan2021knowing}---as a lens for examining the relationships at the heart of the arboretum’s operation. Rather than positioning plants as passive specimens and curators as neutral observers, the focus here is on their mutual entanglement. Trees grow, die, and regenerate within institutional timelines. They are measured, treated, relocated, and occasionally lost. In turn, curators are shaped by these life cycles, developing methods, routines, and forms of care that respond to the needs and rhythms of the plants.

This emphasis on care draws from broader work in STS and feminist science studies, highlighting the affective and ethical dimensions of knowledge production. Puig de la Bellacasa~\cite{ref11}, for example, calls for a shift in attention toward maintenance, repair, and situated care---qualities often overlooked in dominant narratives of scientific progress. Similarly, Jones and Cloke~\cite{ref12} explore how plants and their caretakers co-produce both spatial and emotional geographies. In botanical gardens, these interspecies relations are not peripheral or decorative; they are embedded in the very practices of scientific life: naming, pruning, mapping, and archiving.

\subsection{Archives as Active Interfaces}

Rather than anthropomorphizing plant life, this project emphasizes the activation of archives---approaching data itself as a site of inquiry. As Loukissas~\cite{ref13} notes, botanical data is always local and contingent; its artifacts often reflect the personalities and practices behind their creation. Meaning emerges not in isolation but through the social and institutional frameworks that sustain it, including terminology conventions, database schemas, and visualization tools. In this light, the archive is not a neutral container but a historically layered interface---shaped by the cumulative work of curators, gardeners, and administrators.

The notion of a ``living archive'' thus takes on a double meaning: it points to the biological vitality of the arboretum’s collections and to the evolving data ecosystem that documents them. This project draws on digital humanities approaches that view archives not as static repositories, but as dynamic environments for interpretation, modeling, and recontextualization. This perspective resonates with Bowker and Star’s observation that ``classifications are a key part of the standardization processes that are themselves the cornerstones of working infrastructures''~\cite[p.~231]{ref3}.

The arboretum’s evolving system of plant labels, accession numbers, and inspection records exemplifies this logic---forming a dense, historically layered infrastructure that not only describes but also structures the practices of ecological care. Through computational techniques such as embedding-based clustering, temporal graph analysis, and geospatial plotting, this study surfaces the tacit labor and relational knowledge embedded in these classificatory systems, illuminating how the archive both reflects and enacts institutional memory.

\subsection{Toward a Poetics of Botanical Data}

Finally, computational analysis is paired with aesthetic and narrative modes of engagement. Inspired by projects such as Columbia’s Advanced Data Visualization Project~\cite{advp2025} and the Weather Map~\cite{ref16}, this framework extends visualization beyond representation, drawing on Johanna Drucker’s concept of \textit{capta}---data that are taken, not simply given~\cite{ref5}. The histories of trees, curators, and ecological stewardship are inscribed in the arboretum’s archive, yet they resist being flattened into metrics. The aim, then, is not only to analyze but to render these actors perceptible: as an ecology of care, a choreography of maintenance, and a spatially grounded, temporally extended system of meaning-making.

The research framework brings together ethnobotanical attention to multispecies relations~\cite{ref4}, STS perspectives on infrastructural labor~\cite{ref3,ref6}, and digital humanities techniques for reanimating historical data~\cite{ref5}. The arboretum is more than a garden or a database; it is a space where care, classification, and collaboration meet---across species, disciplines, and generations.


\section{Traces, Layers, and the Living Archive}

Why return to the botanical garden today---not merely as a green space or scientific resource, but as a record of interwoven lives, both human and nonhuman? This section examines the Arnold Arboretum as an exceptional case of archival continuity and infrastructural experimentation. Through its curatorial routines, architectural form, and scientific mission, the arboretum has long functioned as a designed landscape of data---a site where plants, knowledge, and care are cultivated together. Tracing its evolution reveals how historical documents, institutional practices, and ecological systems cohere into a living archive.

\subsection{A Designed Landscape of Data}

The Arnold Arboretum was established in 1872 as a hybrid institution---at once scientific and civic, public and private. Founded through a bequest from James Arnold and administered by Harvard University under a 1{,}000-year lease from the City of Boston, the arboretum represented a pioneering fusion of academic botany and landscape architecture~\cite{ref15}. It was not merely a garden, but a purpose-built research environment, committed to the long-term observation, classification, and cultivation of woody plants.

Charles Sprague Sargent, the arboretum’s first director, envisioned it as a ``scientific station for the study and cultivation of trees''~\cite{ref14}. Under his leadership, the institution implemented one of the first accession-based cataloging systems in the United States---assigning each plant a unique identifier and recording its growth, health, and movement over time. This accession system transformed the landscape into a relational database in the open air, where each tree functioned as both a biological specimen and a node in a wider network of curatorial decisions~\cite{ref14}.

The physical layout was co-designed by Frederick Law Olmsted, whose paths, vistas, and planting schemes reflected not only aesthetic ideals but also practical concerns---such as access, drainage, and microclimate. Olmsted’s design made the arboretum navigable, maintainable, and legible, not just to visitors but to botanists, horticulturists, and public officials. In this sense, the arboretum emerged as a computational object \textit{avant la lettre}---a spatial system structured for observation, comparison, and data retrieval.

From its inception, the arboretum embedded data logic into its terrain. Pathways enabled tracking; planting schemes echoed taxonomic structure; and physical labels on trees were mirrored in handwritten ledgers and printed maps. Early curatorial labor involved not only planting and pruning but also extensive annotation, linking botanical care with textual inscription. The arboretum thus became a machine for generating structured knowledge about plant life and death---season after season, year after year~\cite{ref14,ref15}.

\begin{figure}[tb]
 \centering
 \includegraphics[width=\linewidth]{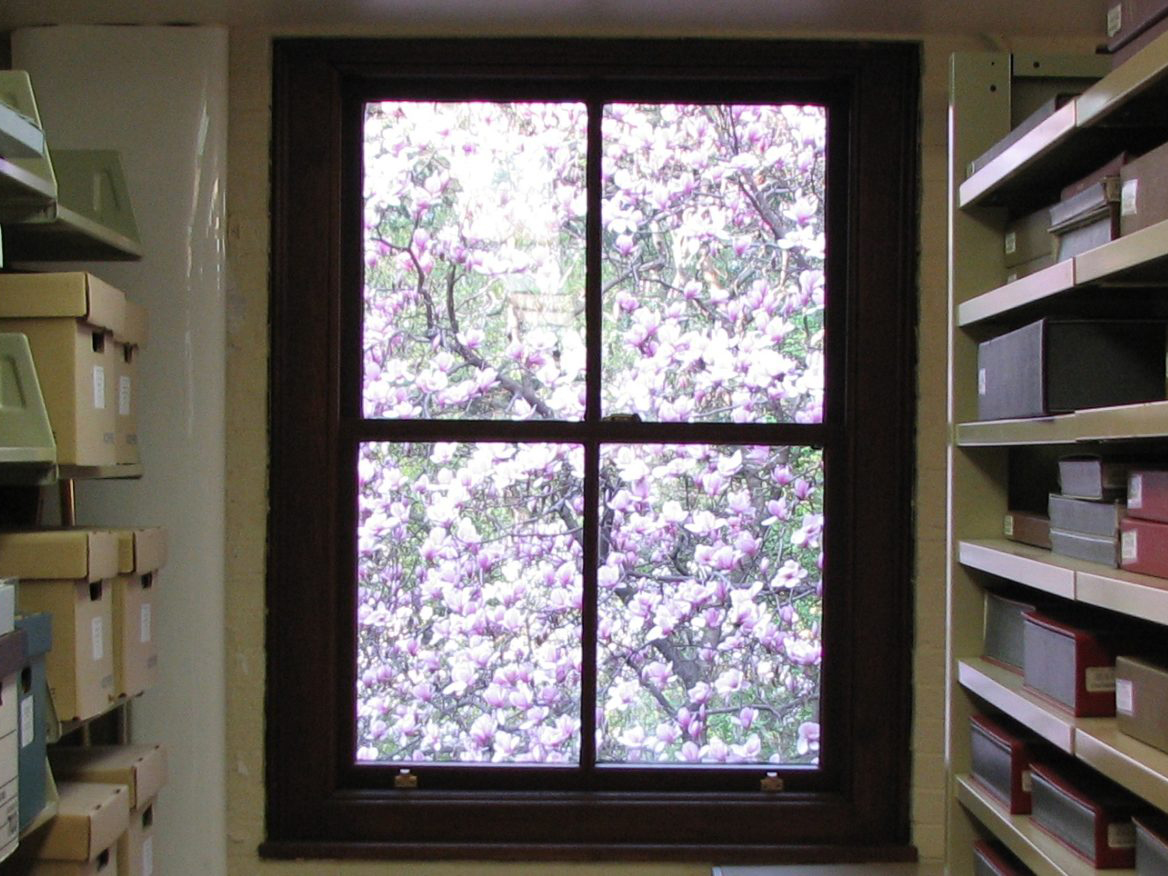}
\caption{View from the Arnold Arboretum’s archives, where shelves of records frame a window onto the living collection, underscoring the link between physical documentation and the landscape it describes [Courtesy of the Arnold Arboretum of Harvard University].} \label{fig:archive}\end{figure}

\begin{figure}[tb]
 \centering
 \includegraphics[width=\linewidth]{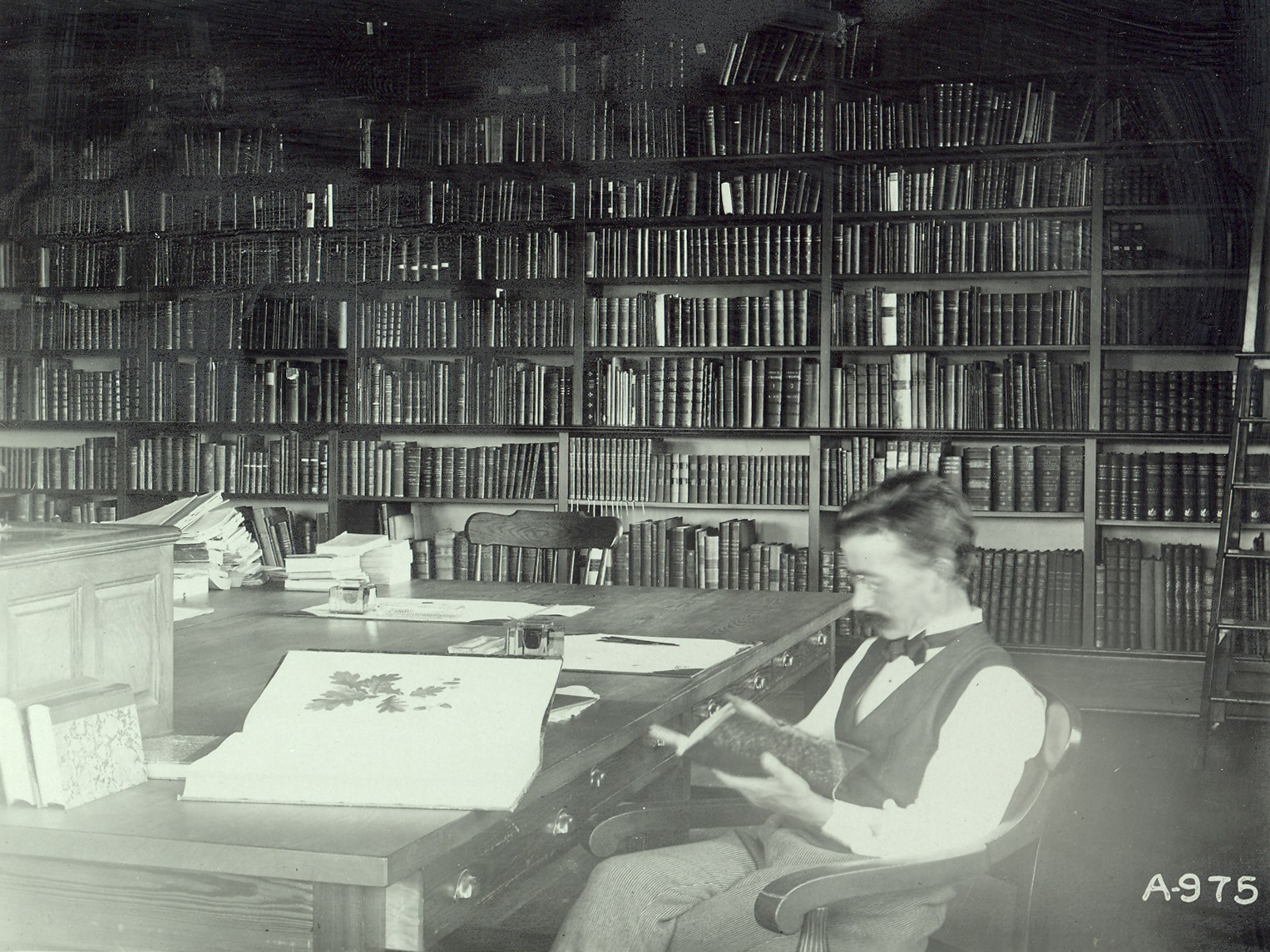}
\caption{Archival photograph of a curator working with plant specimens in the arboretum’s library, highlighting the skilled labor and meticulous care behind the creation and upkeep of botanical records [Courtesy of the Arnold Arboretum of Harvard University].} \label{fig:caretakers}
\end{figure}

\subsection{Archival Depth and Continuity}

Few ecological institutions can claim such an extensive and uninterrupted paper trail. At the Arnold Arboretum recordkeeping began with nineteenth-century field notes and accession books, later expanding to typed memos, herbarium slips, card indexes, and more recently digital formats. Each transition left behind layers of metadata and interpretive sediment, resulting in a uniquely palimpsestic archive – layered, rewritten, and never fully erased~\cite{ref13}.

The accession system documented more than taxonomic facts. It recorded environmental conditions, transplant histories, signs of disease, and causes of death. Each entry accumulated layers of context: who planted a specimen, who moved it, what replaced it, what survived the winter. Trees were not passive elements in the landscape but subjects of care and study, each with a biography – and often an afterlife in research or breeding programs~\cite{ref13}.

This richness makes the arboretum a rare example of institutional memory rendered in vegetal time. It allows for longitudinal analysis of plant lifespans, curatorial regimes, and even climate responsiveness. More than that, it reveals the evolving cultures of knowledge that shaped these records. A nineteenth-century botanist’s ledger entry differs in tone and structure from a database field created by a twenty-first-century data manager. Each medium leaves its own trace – sometimes conflicting, always contingent.

The physicality of the archive matters too. Much of it still resides in drawers, folders, and cabinets, marked by marginalia, shorthand, and annotations that resist digitization. These analog traces offer glimpses into the tacit practices of past curators – their assumptions, routines, and oversights. By bringing together physical and digital records, this project treats the archive as a multi-format ecology, where informational value lies not only in content but in structure, materiality, and omission.

\subsection{The Arboretum as a Knowledge System}

What makes the Arnold Arboretum more than a well-kept repository of plants is its role as a historical knowledge system—one that encodes the values, priorities, and epistemologies of its time. As Tuers~\cite{ref1} and Loukissas~\cite{ref13} observe, botanical data is neither natural nor neutral. Decisions about what to document, how to structure it, and where to store it reflect institutional priorities and infrastructural constraints.

In Tuers’s terms, the arboretum operates as a “planted catalog,” positioned between locality, where specimens are observed and cared for, and the herbarium, where they are abstracted into scientific taxonomy. This transformative border space—where living matter becomes data and care becomes classification—offers a rare opportunity to study botanical knowledge in formation, not at the endpoint of a published monograph but in the daily practices of maintenance and documentation (see \cref{fig:archive} and \cref{fig:caretakers}).

The site also bridges cultural and natural epistemologies. It is simultaneously a garden, a park, a laboratory, and a museum. It is public and private, rooted in local soil yet connected to global networks of plant exchange. In this way, it unsettles binaries between nature and culture, objectivity and intimacy, past and present. It is not a passive repository but an active system—one that must be continually interpreted as well as preserved.

Despite this richness, botanical gardens remain underexamined as  data infrastructures. They are often treated as ancillary to herbaria or museums rather than complex institutions in their own right. By foregrounding the Arnold Arboretum’s combination of institutional continuity, spatial design, and archival practice, this project contributes to a broader rethinking of how living collections serve as laboratories—not only for ecological science but also for multispecies knowledge-making across generations.

\begin{figure*}
 \centering
 \includegraphics[width=\linewidth]{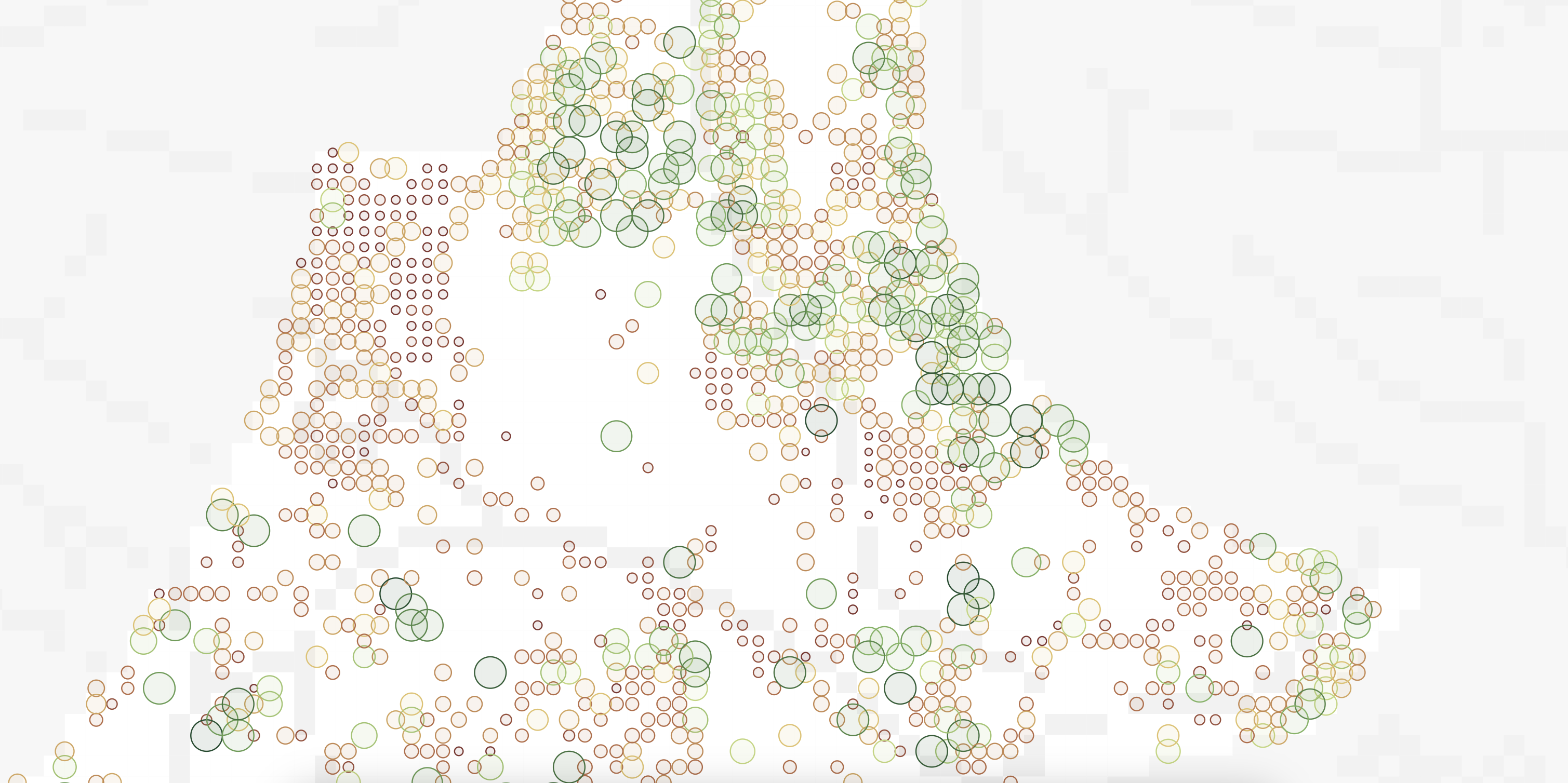}
\caption{Zoomed-out view of the arboretum showing all active notes. The map serves as both a spatial index and an introduction to the project, highlighting areas of concentrated curatorial activity alongside quieter zones. The decades associated with each note, ranging from dark green for the 1870s to deep red for the 2010s and grey for unknown dates, follow the project’s color scale.} \label{fig:map_overview}
\end{figure*}


\section{Methodology: Designing a Visual Interface}

This project is concerned not only with what is shown, but with the conditions, sources, and interpretive stakes that shape its visualization. One aim is to support future and comparative work with botanical collections, which calls for methodological transparency. This section outlines the technical, curatorial, and conceptual choices that guided the development of an interactive interface for the Arnold Arboretum’s living archive.

The resulting platform is not a direct simulation but a discursive reflection on the arboretum. It is a visual narrative that foregrounds the sites where data has been produced, curated, and interpreted, both in the landscape and in the archive. It is as much a representation of institutional and human care as it is a cartography of trees.

\subsection{Data Discovery and Preparation}

The starting point was not the creation of digital records, but their interpretation. The Arnold Arboretum had already undertaken extensive digitization: accession ledgers, planting maps, spreadsheets, curator notes, and other internal documents had been scanned or transcribed into structured and semi-structured formats. This existing digital corpus provided a valuable foundation, enabling large-scale analysis and visualization without the need for preliminary archival processing. Yet, as with any digitized archive, it brought its own set of challenges.

Digitization preserves, but it also abstracts. Artifacts of analog recordkeeping—handwriting quirks, overlapping notations, nonstandard abbreviations—are often flattened in translation. Some fields were inconsistently encoded, some documents loosely structured, and historical idiosyncrasies carried over into digital form. For example, dates might refer to planting, inspection, removal, or death without explicit labels. Curator names appeared in multiple variants (e.g., “J. Malmstedt,” “Johan M.”), and even key identifiers like accession numbers were not always applied consistently.

To navigate this complexity, the focus was placed on entity resolution: writing scripts to harmonize name variants, standardize date fields, and align tree records across time. Where gaps or ambiguities emerged, internal databases and curator records were cross-referenced to recover missing links. This work was less about generating new data than about tracing the epistemic residue of past data practices—a process of reconstruction and reinterpretation that underscores the archive’s layered and contingent nature. In keeping with Loukissas’s view of data as local and contingent~\cite{ref13}, these irregularities were treated not simply as errors to be fixed, but as historical artifacts revealing shifts in curatorial practice, data entry conventions, and institutional priorities.

During this phase, the zoomed-out map view (\cref{fig:map_overview}) became an analytical tool in its own right. Plotting all active notes simultaneously exposed unexpected spatial clusters, temporal gaps, and outliers in the dataset—patterns that pointed to underlying archival inconsistencies. Investigating these anomalies led back to previous storage forms, and even dialogues with employees at the arboretum, reinforcing the idea that the planted catalog~\cite{ref1} exists across multiple media and contexts. The iterative process of plotting, reviewing, and refining not only improved dataset coherence but also aligned the preparation work with the project’s broader aim: to make visible the layered nature of the arboretum’s living archive and to set the stage for its activation in the visual interface.

\subsection{Data Cleaning and Structuring}

Transforming the dataset into a usable form required more than standardization – it demanded close reading of how the data had been entered and maintained over time. The digitized records followed no fixed schema; instead, they reflected evolving curatorial routines. By analyzing fields such as \texttt{CHECK\_BY} annotations and other routine notes, it became possible to identify patterns of institutional labor – who inspected which trees, when, and how consistently. These behaviors, rather than any predetermined structure, formed the foundation for parsing and reorganizing the dataset.

A schema was gradually developed to accommodate overlapping temporalities – planting, inspection, removal – and to represent multiple actors, including plants, curators, and institutional systems. Standardization efforts focused on harmonizing date formats, taxonomic names, curator signature fields, and geospatial coordinates to create a more consistent dataset without erasing historical complexity. In keeping with Bowker and Star’s argument that classification systems actively shape the domains they describe~\cite{ref3}, these standardization choices were understood not as neutral technical steps, but as interpretive interventions that reveal the institutional priorities embedded in the data.

\begin{figure}[tb]
 \centering
 \includegraphics[width=\linewidth]{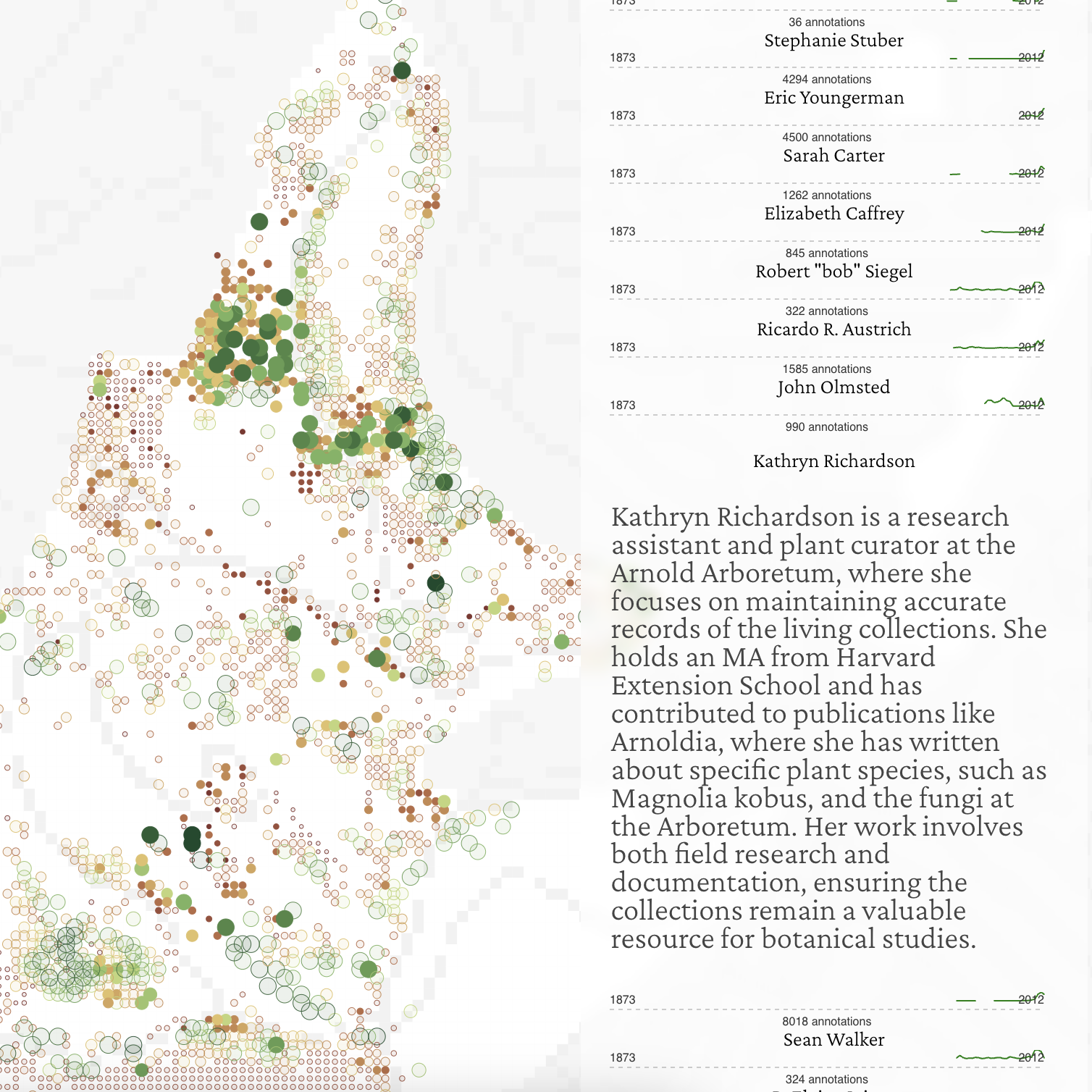}
 \caption{Kathryn Richardson’s work focused on a few sections of the arboretum, with dense clusters of annotations recorded over a relatively short span. Her activity reflects targeted curatorial assignments rather than continuous, long-term coverage.}
 \label{fig:curator_1}
\end{figure}

\begin{figure}[tb]
 \centering
 \includegraphics[width=\linewidth]{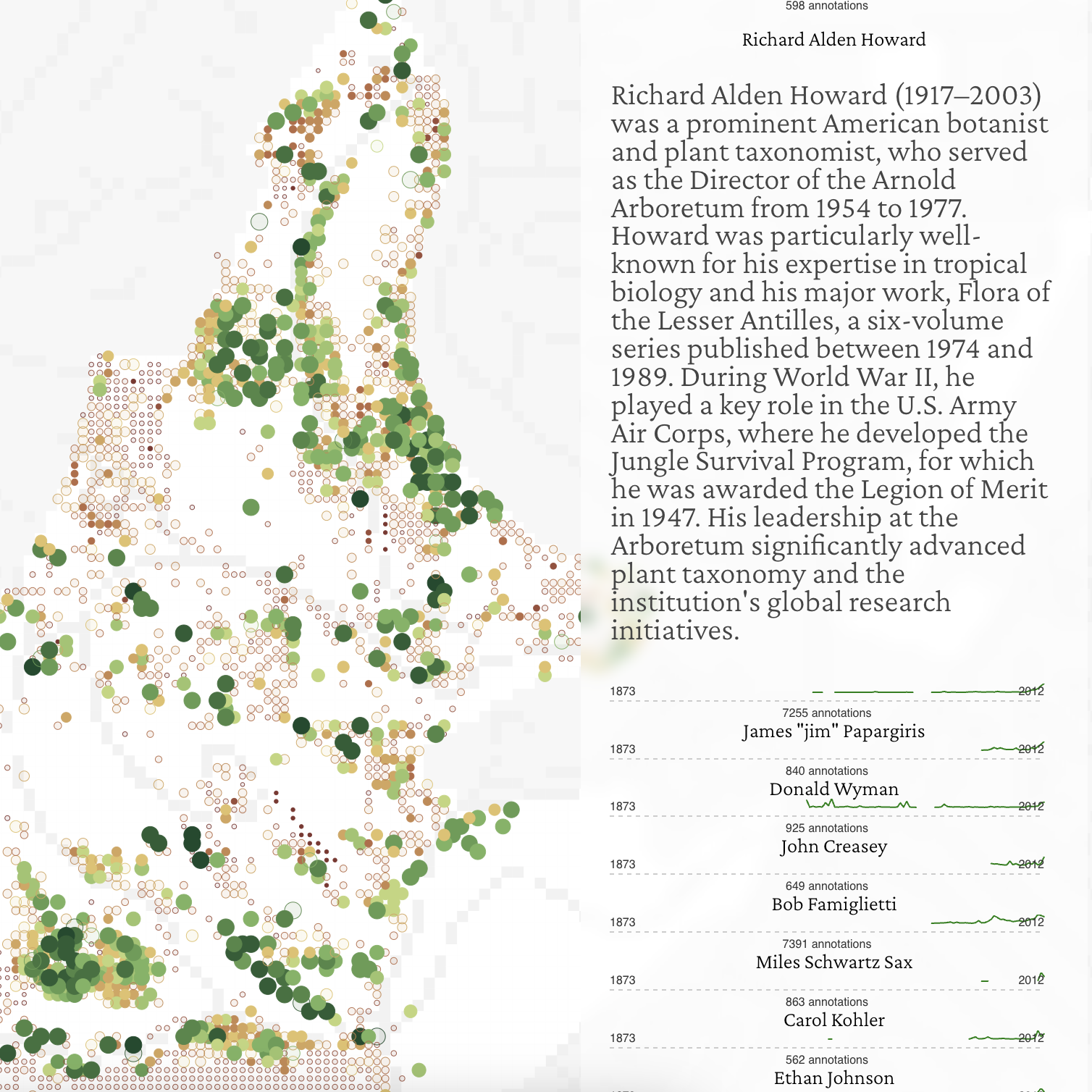}
 \caption{Richard Alden Howard’s records cover a much broader portion of the arboretum and extend across decades, corresponding to his tenure as director from 1954 to 1977. His annotations indicate engagement with both the breadth and continuity of the arboretum.}
 \label{fig:curator_2}
\end{figure}

A key realization emerged during this process: the digital archive does not directly mirror the arboretum’s physical layout. Some trees exist only in annotations; others are extensively documented but no longer present on site. Rather than mapping physical presence, the interface is designed to visualize the imprint of data work—where curatorial attention has left a trace. This principle is most clearly expressed in the Curator Profile Views, which combine harmonized geospatial and temporal data to map each curator’s recorded activities. Alongside these visualizations, narrative biographies situate individual contributions within the arboretum’s broader ecology of care, pairing spatial footprints with personal and professional context. Kathryn Richardson’s records show short-term work in specific sections of the arboretum, while Richard Alden Howard’s span decades and cover a broad range of locations, reflecting his long-term involvement (see \cref{fig:curator_1} and \cref{fig:curator_2}).

Fields like “last check date” or “checked by,” in the original data, became central not for their accuracy or completeness, but for the stories they tell about the rhythms of observation and care. By plotting these attributes visually, patterns emerge that align with the project’s theoretical framing of botanical gardens as knowledge infrastructures~\cite{ref8,ref9,ref10}: spaces where the visibility of certain plants, places, and practices is a direct consequence of historically situated attention. In this way, the visualization positions the arboretum not just as a living collection, but as a dynamic archive of institutional memory, shaped by the shifting interactions between people, plants, and data systems.

\subsection{Machine Learning and Enrichment}

Given the dataset’s uneven coverage, language models were used as interpretive tools to generate speculative biographies of both trees and curators—narrative sketches that blend structured archival fields with unstructured textual sources, including digitized ledger notes, website entries, and articles from \emph{Arnoldia}, the institution’s long-running journal.

Using large language models, prompts were crafted to synthesize contextual summaries, particularly when records were sparse or encoded in shorthand. For instance, entries like “WDS, TM ’99” were parsed and rephrased using known personnel data and recognizable curatorial patterns. These summaries were not meant to resolve ambiguity but to make its contours legible. In this sense, the language models were mobilized in the spirit of Drucker’s concept of \emph{capta}~\cite{ref5}—not as neutral processors of given facts but as interpretive agents producing situated readings of archival traces.

This approach acknowledged the patchiness of archival memory: some trees were meticulously recorded across decades, while others received only fleeting mention. Rather than conceal these gaps, the interface was designed to highlight them. In the curator profiles (see \cref{fig:curator_1} and \cref{fig:curator_2}), AI-generated narrative fragments are embedded directly alongside each curator’s spatial and temporal activity maps, providing context for their recorded footprints. This pairing ensures that speculative interpretation and empirical record remain visible side by side, enabling the user to weigh both in forming their own understanding.

By integrating these enriched narratives into the visualization, the system enacts the project’s broader goal of creating a poetics of botanical data~\cite{ref5,ref16}. The AI outputs do not close interpretive questions; instead, they open them further, framing uncertainty as a meaningful part of the record. The result is not a definitive history but an interpretive landscape shaped by the textures of curatorial care and omission, resonating with the notion of the archive as an active interface~\cite{ref13}—a place where human judgment, technical process, and more-than-human histories converge in visible form.

\begin{figure}[tb]
 \centering
 \includegraphics[width=\linewidth]{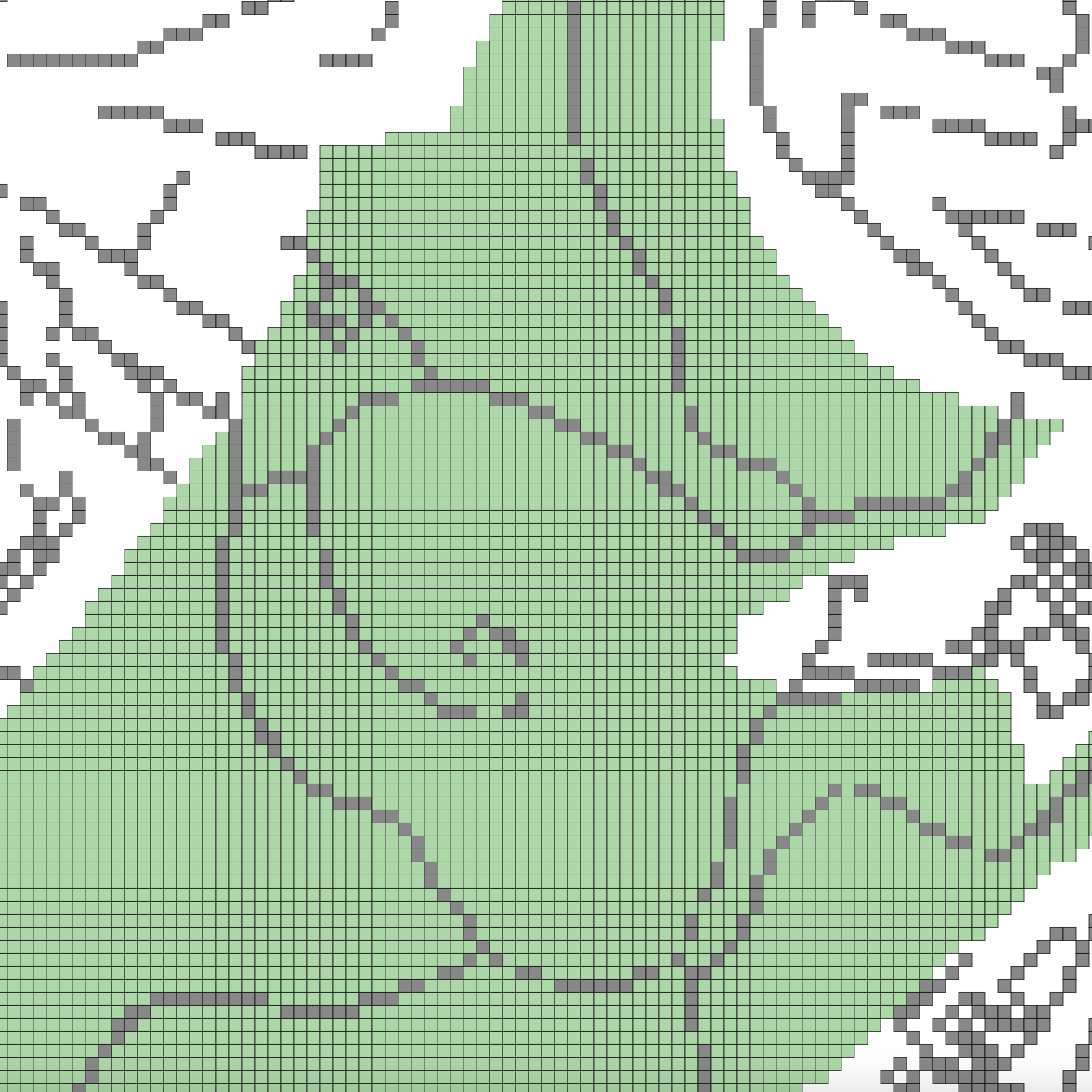}
 \caption{Early stage of the grid-based discretization process with a fine resolution (grid size = 5), capturing detailed spatial features of the arboretum’s topography. This higher density supports greater locational accuracy but increases visual and computational complexity.}
 \label{fig:map_grid_2}
\end{figure}

\begin{figure}[tb]
 \centering
 \includegraphics[width=\linewidth]{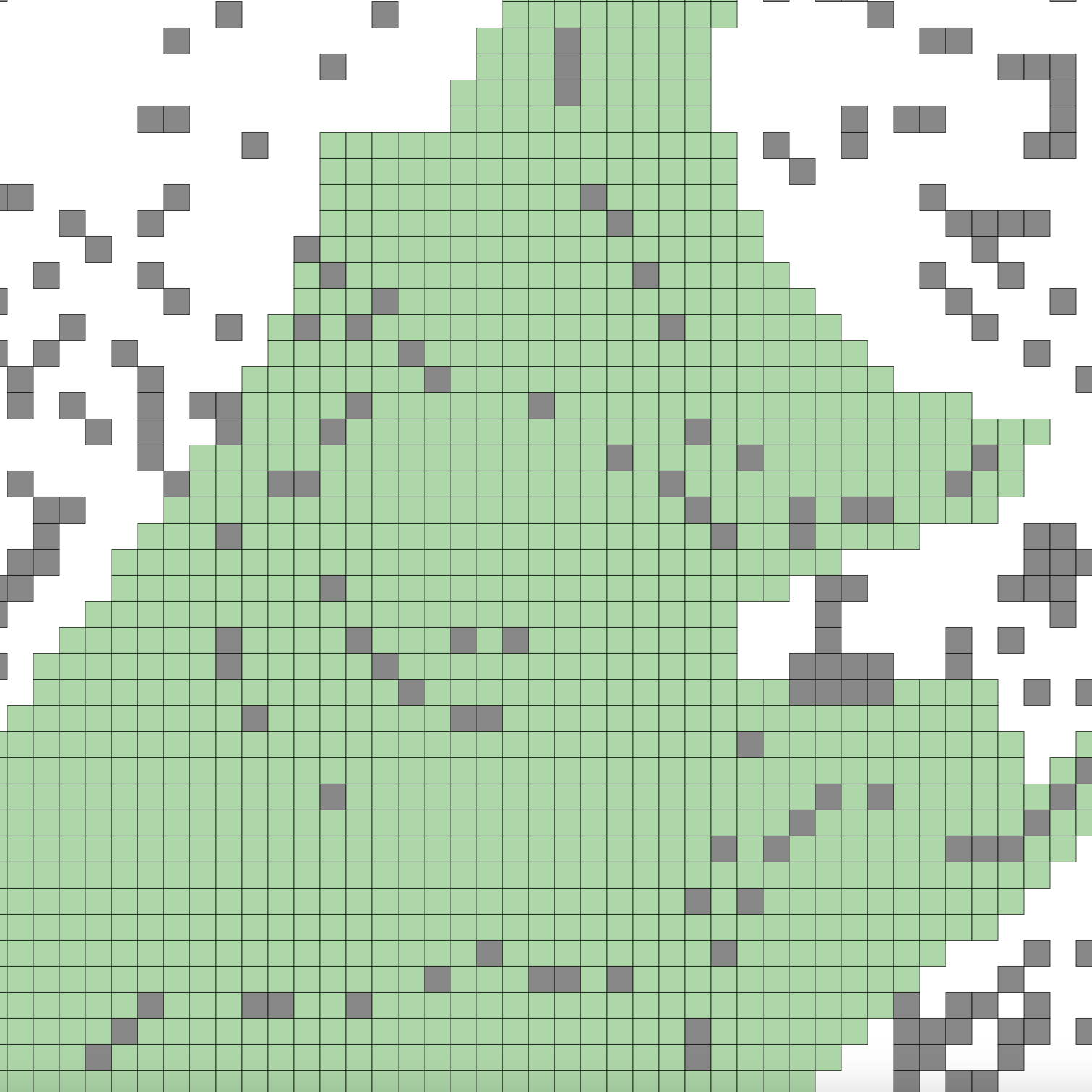}
 \caption{Coarser discretization of the Arnold Arboretum’s map (grid size = 20), emphasizing broad spatial patterns over fine-grained detail. This reduced density streamlines computation and highlights general structures at the expense of precision.}
 \label{fig:map_grid_1}
\end{figure}

\subsection{Design and Interaction}

Early in the process, we experimented with grid-based discretization to translate the arboretum’s topography into a computationally manageable form. Fine-resolution grids (Fig. \ref{fig:map_grid_2}) captured detailed spatial features and supported greater locational accuracy, but at the cost of higher visual and computational complexity. Coarser grids (Fig. \ref{fig:map_grid_1}) emphasized broader spatial patterns and streamlined processing, though they sacrificed precision. These trials informed the choice of a resolution that balanced spatial fidelity with clarity, and they also established an early design principle: that spatial representation should privilege interpretability and narrative potential over exhaustive precision. This foundation proved critical as the project shifted toward more conceptually driven visual forms.

\begin{figure*}[tb]
  \centering
  \includegraphics[width=\textwidth]{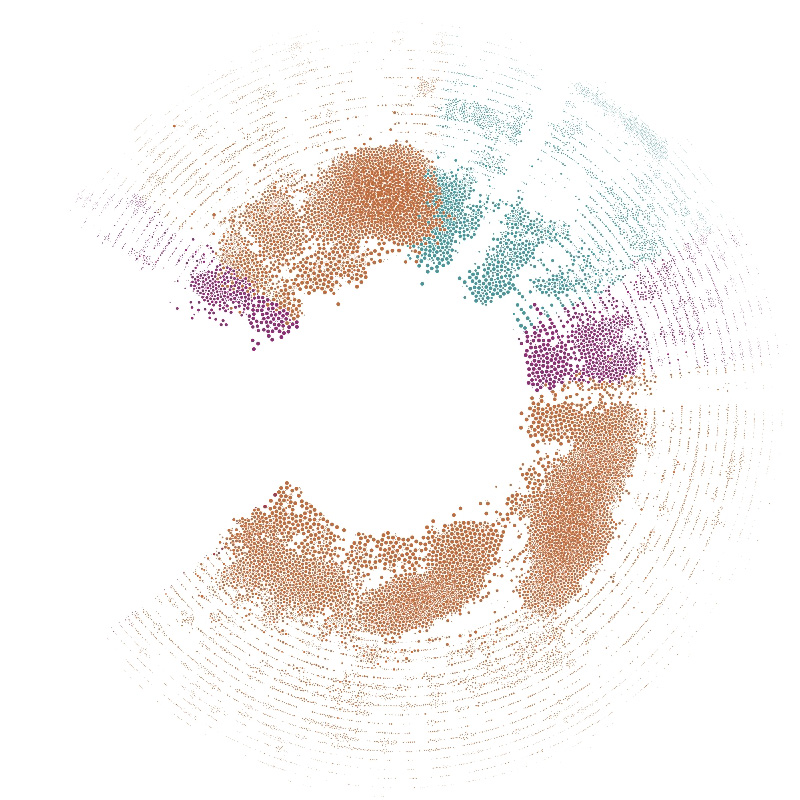}
  \caption{An early interface prototype using a tree-ring metaphor to map the temporal and relational dimensions of the arboretum’s archive. Each concentric ring represents a year in the institution’s history, while individual marks record the presence and activity of specific trees or curators. The form draws on the natural growth patterns of tree rings to convey the passage of time and the accumulation of care, linking ecological cycles with archival processes. Although later iterations shifted toward a map-based structure, the tree-ring design informed the platform’s overall visual language and its emphasis on layered temporal narratives.}
  \label{fig:tree_ring}
\end{figure*}

From the outset, the goal was not to build a dashboard but to create a narratively layered encounter with the arboretum’s archival ecosystem. This approach reflects the project’s framing of botanical gardens as knowledge infrastructures \cite{ref8,ref9,ref10}, where spatial and temporal patterns are inseparable from the histories of their documentation. Early visual metaphors explored ways to connect the temporal depth of the archive with the spatial logic of the site. One early sketch (see \cref{fig:tree_ring}) centered on the motif of a tree ring---a visual form evoking growth, time, and memory. In this prototype, the arboretum’s history unfolded as concentric rings: each year a layer, each tree a trace. This design was not merely aesthetic; it aimed to link the life cycles of the trees to the accumulation of curatorial records, making time visible as a layered material in its own right.

While the map view eventually became the primary organizing structure—favored for its clarity, density of information, and ability to reveal patterns at multiple scales---the tree-ring metaphor continued to inform the interface’s visual philosophy. Radial timelines and animated growth patterns carried forward the conceptual link between natural cycles and archival accumulation. The resulting visual language allowed the interface to signal temporal change without overwhelming the user, balancing static clarity with dynamic cues that encouraged sustained exploration.

Design choices were shaped equally by technical constraints and interpretive aims. Some arose from data gaps---for example, the absence of consistent curator biographies prompted the development of exploratory narrative pathways rather than a fixed chronological view. Others stemmed from a desire to create a particular atmosphere: color gradients were tuned to echo the palette of the arboretum across seasons; subtle animations mimicked the pacing of walking through a garden; and spatial composition was arranged to invite unhurried exploration rather than rapid extraction of facts. The interface was designed not just to convey information, but to slow the user down, encouraging reflection on the rhythms of care embedded in the archive.

The final interface supports layered engagement, offering different points of entry for arboretum staff, academic researchers, and the general public. Metadata are presented alongside interpretive texts and open-ended visualizations, enabling users to approach the archive through factual detail, narrative framing, or aesthetic experience. This multiplicity reflects the arboretum’s own identity as both a scientific and public institution, where the same space can be a site for scholarly analysis, casual wandering, and historical reflection. In this way, the design bridges disciplinary divides, inviting users to see the living collection not only as a dataset to be mined but as a cultural and ecological record to be experienced.

The project remains ongoing, with further features and datasets in development. The source code and the current version of the application are available at \url{https://sinanatra.github.io/arboretum/}, supporting reuse by researchers, curators, and the public alike~\cite{arboretum_github}. Developed primarily in JavaScript, the application draws on a range of web libraries such as Svelte and D3.js to ensure responsiveness, interactivity, and ease of deployment. Additional libraries—such as TopoJSON for compact geographical data handling, and Leaflet for lightweight interactive mapping—extend the application’s ability to integrate spatial datasets and visual layers. Together, these tools allow the application to run efficiently in a browser environment without requiring server-side processing.

\clearpage 


\section{Conclusion}

Developing this interface revealed a productive tension between archival depth and computational design. Many decisions arose not from ideal conditions, but from the constraints and gaps embedded in the data. At times, visualization softened those gaps; at others, it brought them into sharper relief. In this way, the interface does not seek to resolve archival complexity—it invites engagement with it. As Johanna Drucker might suggest~\cite{ref5}, it is a method for thinking with the archive, using visual form not as an endpoint but as an instrument for interpretation.

Several insights emerged in the process. The consistent annotations of certain curators, maintained over decades, highlighted care as an ongoing and situated practice, more than a technical duty~\cite{ref11}. In contrast, stretches with little or no metadata pointed to shifting institutional priorities, changes in staffing, or moments of fatigue. The data, like the garden it describes, follow its own rhythms of attention, revealing the interplay between human labor and the temporalities of the living collection. These patterns suggest that botanical archives can serve not only as repositories of taxonomic information but also as records of institutional culture and working memory.

Humanities reasoning proved indispensable throughout. It guided the reading of ambiguous entries, the visual emphasis on absence, and the design of an interface that privileges meaning over utility. The work affirms Tuers’s idea of the “planted catalog” as a layered space where locality, labor, and metadata converge~\cite{ref1}, and echoes Haraway’s argument that knowledge is always situated and co-produced~\cite{ref9,ref17}—here, through the entanglement of plants, people, and paper. In framing the arboretum as a knowledge infrastructure, the project contributes to broader conversations in science and technology studies about how care, classification, and ecological stewardship can be made legible through design.

Ultimately, the project rests on a dual ambition: to honor the specificity and complexity of the Arnold Arboretum’s historical records, and to model how digital humanities can approach botanical collections not merely as archives of knowledge, but as infrastructures for co-creating meaning across time, species, and disciplines~\cite{ref13,ref18}. 

\section*{Acknowledgments}

We thank metaLAB at Harvard for providing an intellectual and collaborative home for the work, and in particular its three directors—Jeffrey Schnapp, Annette Jael Lehmann, and Aylin Yildirim Tschoepe—for their encouragement and support throughout the project. We are also grateful to the Arnold Arboretum of Harvard University for granting access to its data and image collections, which form the foundation of this research, and to Matthew Battles, former associate director of metaLAB and now Editor of \textit{Arnoldia}, for his generous insights and support.

\newpage

\bibliographystyle{abbrv-doi}
\bibliography{references}

\end{document}